\documentclass[epj,twocolumn]{webofc}
\usepackage[varg]{txfonts}   
\woctitle{CGS15}
\begin{document}
\title{Measurements of neutron-induced reactions in inverse kinematics and applications to nuclear astrophysics}
\author{Ren\'e Reifarth\inst{1}\fnsep\thanks{\email{reifarth@physik.uni-frankfurt.de}}  \and 
Yuri A. Litvinov\inst{2} \and 
Anne Endres\inst{1} \and 
Kathrin G\"obel\inst{1} \and 
Tanja Heftrich\inst{1} \and 
Jan Glorius\inst{1,2} \and
Alexander Koloczek\inst{1,2} \and 
Kerstin Sonnabend\inst{1}  \and
Claudia Travaglio\inst{3,4}  \and
Mario Weigand\inst{1} 
}

\institute{Goethe University Frankfurt am Main, Frankfurt, Germany \and
GSI Helmholtzzentrum für Schwerionenforschung, Darmstadt, Germany \and
INAF-Astronomical Observatory Turin, Turin, Italy \and
B2FH Association, Turin, Italy
}

\abstract{Neutron capture cross sections of unstable isotopes are important for neutron-induced nucleosynthesis as well as for technological applications. A combination of a radioactive beam facility, an ion storage ring and a high flux reactor would allow a direct measurement of neutron induced reactions over a wide energy range on isotopes with half lives down to minutes.
The idea is to measure neutron-induced reactions on radioactive ions in inverse kinematics. This means, the radioactive ions will pass through a neutron target. In order to efficiently use the rare nuclides as well as to enhance the luminosity, the exotic nuclides can be stored in an ion storage ring. The neutron target can be the core of a research reactor, where one of the central fuel elements is replaced by the evacuated beam pipe of the storage ring. Using particle detectors and Schottky spectroscopy, most of the important neutron-induced reactions, such as 
(n,$\gamma$), (n,p), (n,$\alpha$), (n,2n), or (n,f), could be investigated.}
\maketitle
\section{Introduction}
\label{intro}
\begin{figure}[h]
\centering
\includegraphics[width=0.495\textwidth]{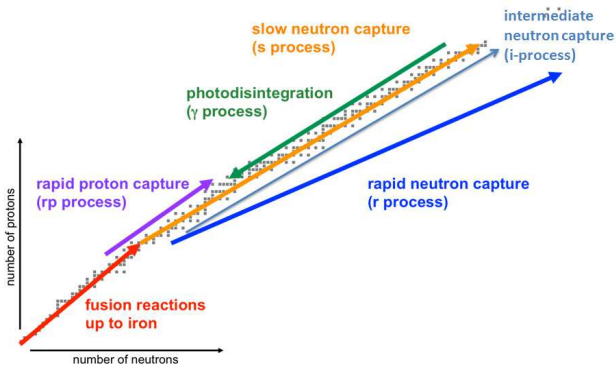}
\caption{The $s$ and $r$ processes start with the
  iron peak nuclei as seeds.
  The s process path follows the nuclear valley of stability until it terminates
  in the lead-bismuth region.
  The r process drives the nuclear matter far to the neutron-rich side of the
  stability line and upwards until
  beta-delayed fission and neutron-induced fission occur and recycle the material
  back to smaller mass numbers. The reaction path of the $i$ process lays in between, since the corresponding
  intermediate neutron densities are higher than during the $s$ process, but still much smaller 
  than during the $r$ process.
  Only a few isotopes on the proton-rich side of the valley of stability get
  significant contributions from the different models of the $p$ process.
\label{reifarth_fig_synthesis_elements}}
\end{figure}

Almost all of the heavy elements are produced via neutron capture reactions equally shared between
$s$ and $r$ process \cite{AnG89, TFH07, RLK14} and to a very small extent by the $i$ process \cite{HPW11, GZY13}. 
The remaining minor part is produced
via photon- and proton-induced reactions during the $p$ process, see Fig.~\ref{reifarth_fig_synthesis_elements}. 
The predictive power of the underlying 
stellar models is limited because they contain poorly constrained 
physics components such as convection, rotation or magnetic fields. The crucial link between the observed abundances 
\cite{Lod03} and the desired parameters of the stellar interiors are nuclear
reaction data. In contrast to the $r$ process \cite{TMP07}, the isotopes important for
the $p$ process \cite{ArG03} as well as for the $s$ and $i$ process \cite{KGB11} are stable or not too far off the valley of stability.
The nuclear properties of those nuclei are therefore experimentally much easier to access. 
While indirect measurements, like the investigation of time-reversed reactions, are very often 
the only choice because of the short half-lives of the investigated isotopes \cite{LLA14,AAA14,GAA14},
this article is focussing
on experimental developments aiming at a better understanding of the $s$, $i$, and $p$ processes by direct measurements
of important neutron capture reactions. 

Neutron capture cross sections of stable and unstable isotopes 
are important 
for neutron-induced nucleosynthesis \cite{RLK14} as well as for technological 
applications \cite{CHO11}. Following a review of different astrophysical nucleosynthesis 
sites with an emphasis on neutron-induced reactions, we will present the capabilities of a 
combination of a radioactive beam facility, an 
ion storage ring and a high flux reactor. Such a facility would allow a direct measurement of 
neutron induced reactions over a wide energy range on isotopes with half lives 
down to minutes \cite{ReL14}.

\section{Astrophysical Scenarios}
\subsection{$p$ process}
The most intensively
studied astrophysical site of the $p$ process is the O-Ne layer of a massive star passed by the shock wave of a
type II supernova \cite{HIK08}. Here, the $p$ nuclei are produced in a series of photodisintegration 
reactions, such as ($\gamma$,n), ($\gamma$,p), and ($\gamma$,$\alpha$) reactions, from a given seed abundance. This picture is
often referred to as $\gamma$~process \cite{ArG03}. The reaction network to describe this scenario includes a
huge number of reaction rates, which are generally calculated in the framework of the 
Hauser-Feshbach statistical model.

An alternative model are supernovae Ia, which are associated with thermonuclear explosions of white dwarfs.
An often proposed scenario is an explosion triggered once the star approaches
the Chandrasekhar mass because of mass accretion from a companion main-sequence star. 
This is referred to as the single-degenerate scenario~\cite{HiN00}.

\begin{figure}[h]
\centering{
\includegraphics[width=0.3\textwidth]{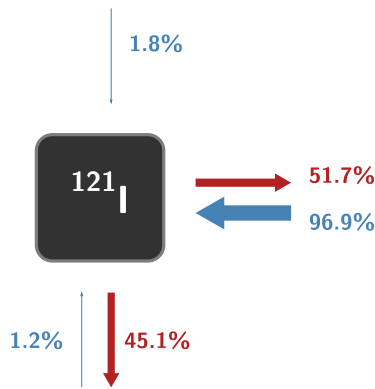}
\caption{Reaction flows for $^{121}$I during the $\gamma$ process. Blue arrows correspond to production and red arrows correspond to
destruction paths. The most important production channel is $^{122}$I($\gamma$,n), 
while the destruction is almost equally shared between $^{121}$I($\gamma$,p) and $^{121}$I(n,$\gamma$). Neutron capture and proton
removal are competing with each other.
\label{reifarth_fig_121i}
}
}
\end{figure}

\begin{figure}[h]
\begin{flushright}
	\includegraphics[width=0.3\textwidth]{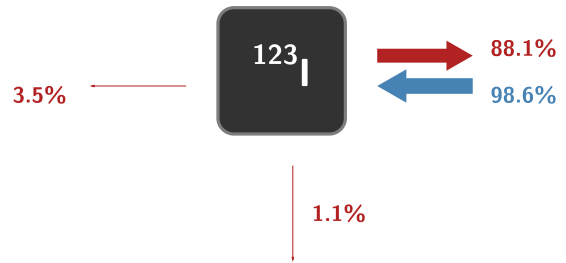}	
\end{flushright}
\caption{Reaction flows for $^{123}$I during the $\gamma$ process. Blue arrows correspond to production and red arrows correspond to
destruction paths. The most important production channel is $^{124}$I($\gamma$,n), 
and the destruction is dominated by $^{123}$I(n,$\gamma$).The (n,$\gamma$)-($\gamma$,n) equilibrium is established.
\label{reifarth_fig_123i}}
\end{figure}

Figures~\ref{reifarth_fig_121i} and \ref{reifarth_fig_123i} show the results of nucleosynthesis
calculations following tracer particles during a supernova type Ia explosion \cite{TRG11,BTG14}. The production and destruction mechanisms for the
neutron-deficient isotopes $^{121,123}$I have been investigated within the framework of the Nucleosynthesis Grid (NuGrid) research platform~\cite{PiH12}. Under the conditions investigated, the neutron evaporation 
process stalls at those isotopes leading to a (n,$\gamma$)-($\gamma$,n) equilibrium as already suggested by \cite{ArG03}. This
means, neutron capture cross section for neutron-deficient isotopes are important for the understanding of the $\gamma$ process.

\subsection{s process}

The modern picture of the main $s$-process component refers to the He shell burning phase 
in AGB stars \cite{LHL03}. Nuclei with masses between 90 and 209 are mainly produced during
the main component. The highest neutron densities in this model occur during the 
$^{22}$Ne($\alpha$,n) phase and are up to 10$^{12}$~cm$^{-3}$ with temperatures around $kT~=~30$~keV.
The other extreme can be found during the $^{13}$C($\alpha$,n) phase where neutron densities
as low as 10$^{7}$~cm$^{-3}$ and temperatures around $kT~=~5$~keV are possible. 
Similarly to the main component, also the weak component referring to different evolutionary stages
in massive stars has two phases \cite{TGA04,HKU07}. Nuclei with masses between 56 and 90 are mainly produced during
the weak component. The first phase occurs during the helium core burning with
neutron densities down to 10$^{6}$~cm$^{-3}$ and temperatures around $kT~=~25$~keV. The second phase
happens during the carbon shell burning with neutron densities up to 10$^{12}$~cm$^{-3}$ at temperatures
around $kT~=~90$~keV.

\begin{figure}[h]
\includegraphics[width=0.495\textwidth]{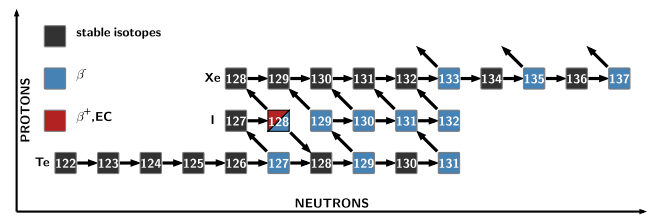}
\caption{The $s$-process nucleosynthesis network around the stable $^{127}$I.
\label{reifarth_fig_reactions_s-process_te-cs}}
\end{figure}

Figure~\ref{reifarth_fig_reactions_s-process_te-cs} shows the nucleosynthesis path of the $s$ process in the region
around iodine. The neutron capture cross sections of $^{127,129,130,131}$I are important in order to understand the nucleosynthesis
of the xenon isotopes in general and in particular the $s$-only isotopes $^{128,130}$Xe \cite{RKV04}.

\subsection{i process}

\begin{figure}
\begin{center}
  \includegraphics[width=.49\textwidth]{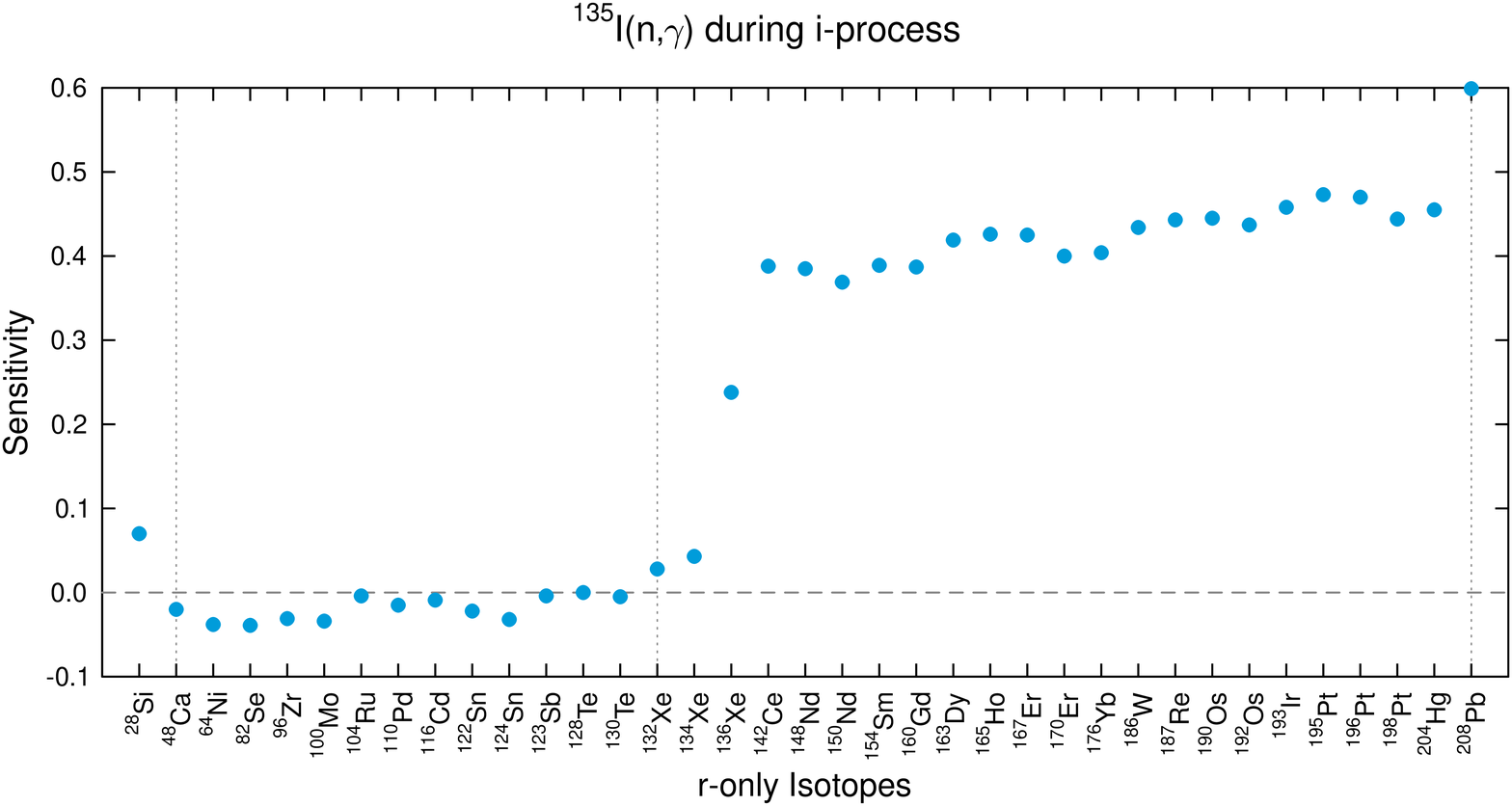}
\end{center}
  \caption{Impact of the $^{135}$I(n,$\gamma$) rate on the final abundances of the i process. 
  This reaction rate affects most of the abundances beyond $^{135}$I and is therefore of global importance. 
  The sensitivity is defined as the ratio between the relative change in abundance and the relative change of the rate.
  \label{reifarth_fig_i_process_I135_sens}}
\end{figure}

Under certain conditions, stars may experience convective-reactive nucleosynthesis episodes. It has been shown with 
hydrodynamic simulations that neutron densities in excess of $10^{15}$~cm$^{-3}$ can be reached \cite{HPW11, GZY13}, if unprocessed, H-rich material is convectively mixed with an He-burning zone. Under such conditions, which are between the $s$ and $r$ process, the reaction flow occurs a few mass units away from the valley of stability. These conditions are sometimes referred to as the i process (intermediate process). One of the most important rates, but extremely difficult to determine, is the neutron capture on $^{135}$I, Figure~\ref{reifarth_fig_i_process_I135_sens}. The half-life of $^{135}$I is about 6~h. Therefore, the $^{135}$I(n,$\gamma$) cross section cannot be measured directly with current facilities. 

\section{Neutron capture measurements in inverse kinematics}

As discussed in the previous section on the example of the isotopic chain of iodine, 
the astrophysical interesting neutron capture rates range from the neutron-deficient
$^{121}$I ($t_{1/2}=2.1$~h) to the neutron-rich $^{135}$I ($t_{1/2}=6.0$~h) and more
long-lived isotopes in between. Ion storage rings turned out to be powerful tools for the investigation
of charged-particle-induced reactions in inverse kinematics \cite{ZAB10,LBB13}. The major advantage 
is the possibility of effectively shooting the ions through a thick target by passing a thin target
multiple times. Since the corresponding cross sections are dominated by the tunneling probability 
through the Coulomb barrier, they show a very strong energy dependence. It is therefore desirable to 
re-accelerate the ions after each pass, which is possible using electron coolers.

\begin{figure}[h]
\centering
\includegraphics[width=0.495\textwidth]{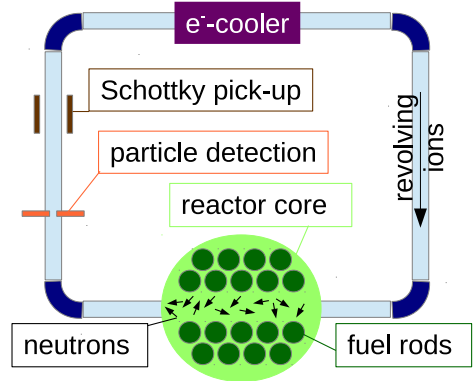}
\caption{Schematic drawing of the proposed setup. 
  Shown are: The main components of an ion storage ring which include the beam lines and focusing elements (blue), 
  dipoles (dark blue), electron cooler (purple), an intersected reactor core (green), 
  particle detection capability (orange) and Schottky pick-up electrodes (brown) \cite{ReL14}.
\label{reifarth_fig_setup_ng_inverse}}
\end{figure}

This idea can be taken a step further by considering a neutron target. Since neutrons are unstable, 
they will have to be constantly produced. This can be done very efficiently using a reactor. The beam
pipe containing the revolving ions has to go through the neutron field, which means, either through or 
at least close by the reactor core, see Figure~\ref{reifarth_fig_setup_ng_inverse}.

In order to discuss the possible reactions, which could be investigated with this setup, it is important to understand the kinematics. The radius $(r)$ of a trajectory of a charged $(q)$ massive $(m)$ particle with velocity $(v)$ in a homogeneous, perpendicular magnetic field $(B)$ follows immediately from the Lorentz force:

\begin{equation}\label{eq_radius_magnetic_field}
  r = \frac{m v}{q B} = \frac{p}{q B}
\end{equation}

Equation \ref{eq_radius_magnetic_field} is even valid  for relativistically moving particles, if $p$ and $m$ are relativistic variables. 
Compared to the revolving beam energy (energies above 0.1~AMeV), the neutrons (energies of 25~meV) can always be considered to be at rest. 
If one assumes that all channels can be viewed as a compound reaction, first a nucleus $X+n$ is formed and in a second step, particles or photons are emitted. This means, the momentum and the charge of the revolving unreacted beam $X$ and the compound nucleus $X+n$ are the same, hence both species will be on the same trajectory. However, the velocity, hence the revolution frequency, is reduced by the factor $A/(A+1)$. If the revolving ions have charge $Z=q/e$ and mass $A=12\cdot m/m_{^{12}C}$, one finds then for the ratio of radii:

\begin{equation}
\frac{r_{D}}{r_{P}} = \frac{Z_P}{Z_{D}}\frac{p_{D}}{p_{P}}=\frac{Z_P}{Z_{D}}\frac{A_P}{A_P+1} \frac{A_{D}}{A_P},
\end{equation}
where indices $D$ and $P$ denote the produced daughter and unreacted parent nuclei, respectively.
And finally one obtains:

\begin{equation}\label{eq_ratio_radius_magnetic_field}
\frac{r_{D}}{r_{P}}=\frac{Z_P}{Z_{D}}\frac{A_{D}}{A_P+1}
\end{equation}

\begin{figure}[h]
\centering
\includegraphics[width=0.495\textwidth]{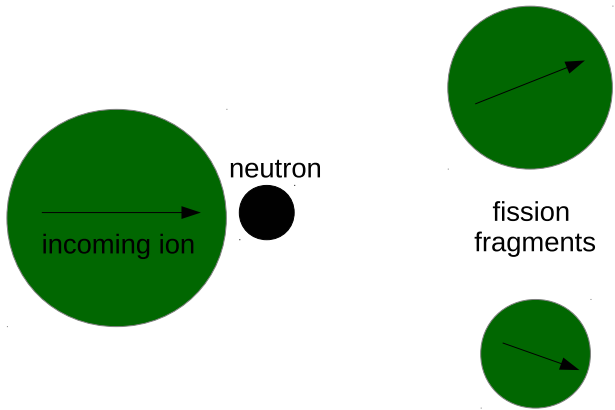}
\caption{Because of the huge amount of energy released during a fission event, the fission products will only stay in a forward cone, if the 
beam energy is above $E\approx10$~AMeV.
\label{reifarth_fig_nf_ng_inverse}}
\end{figure}

It is important to note, that the underlying assumption for these simple equations is that the Q-value of the reaction can be neglected compared to the
energy of the revolving ions. If this assumption is not valid anymore, the trajectories of the products will scatter. This holds true in particular for 
(exothermal) fission reactions, where the Q-value is in the order of 1~AMeV, Figure~\ref{reifarth_fig_nf_ng_inverse}. Therefore the detection mechanisms 
described below will only
be possible for fission, if the beam energy is higher than $\approx10$~AMeV. This corresponds to neutron-induced fission cross section for 
neutron energies of 10~MeV or higher.

If the magnetic rigidities of the reaction products and the primary beam are different, the reaction products can be detected using particle 
detectors placed in the ultra-high ring vacuum. This is for instance possible with double sided silicon strip detectors on ceramic carriers. 
The position and energy deposition in the detector allows the distinction between the reaction products of interest and background \cite{ZAB10}.
According to Eq.~(\ref{eq_ratio_radius_magnetic_field}) this holds true for (n,$\alpha$), (n,p), (n,2n) and (n,f) reactions.

Neutron capture reactions have to be treated differently, since projectile (primary beam) and product will have the same trajectory, 
Eq.~(\ref{eq_ratio_radius_magnetic_field}). In combination with an electron cooler, the number of daughter ions can then be monitored 
by a non-destructive Schottky spectroscopy~\cite{GBB10, NHL11} or by using a sensitive SQUID-based CCC-detectors~\cite{VGN13}. It has been 
shown that even single ions can be detected, even if the primary beam is still present in the ring \cite{SCL13}. 

Storage rings can be moved. Currently there are detailed plans, published as a technical design report, to move the Test Storage Ring (TSR), 
which was in operation until 2013 at the Max-Planck Institute for Nuclear Physics in Heidelberg to CERN where
it shall be coupled to the ISOLDE radioactive ion beam facility~\cite{GLR12}.

An ISOL-type radioactive ion beam facility could be the source of the exotic nuclei. ISOL-beams combine high intensity and good quality~\cite{SKL07}. 
For this particular application, the charge state of the revolving ions is not important, except for the constraints concerning the difference in 
trajectory with charged particles in the exit channel.

\begin{acknowledgement}
This project was supported by the Helmholtz International Center for FAIR, the Helmholtz-CAS Joint Research Group HCJRG-108, the GIF project G-1051-103.7/2009, the BMBF project 05P12RFFN6 and the EuroGenesis project MASCHE. The numerical calculations for tracer particles in SNIa have been supported by the B2Fh Association and the Goethe University Frankfurt.
\end{acknowledgement}

\newcommand{\noopsort}[1]{} \newcommand{\printfirst}[2]{#1}
  \newcommand{\singleletter}[1]{#1} \newcommand{\swithchargs}[2]{#2#1}


\begin{thebibliography}{31}

\bibitem{AnG89}
E.~Anders, N.~Grevesse, Geochim. Cosmochim. Acta \textbf{53}, 197 (1989)

\bibitem{TFH07}
F.K. {Thielemann}, C.~{Fr{\"o}hlich}, R.~{Hirschi}, M.~{Liebend{\"o}rfer},
  I.~{Dillmann}, D.~{Mocelj}, T.~{Rauscher}, G.~{Martinez-Pinedo},
  K.~{Langanke}, K.~{Farouqi} et~al., Progress in Particle and Nuclear Physics
  \textbf{59}, 74 (2007)

\bibitem{RLK14}
R.~{Reifarth}, C.~{Lederer}, F.~{K{\"a}ppeler}, Journal of Physics G Nuclear
  Physics \textbf{41}, 053101 (2014)

\bibitem{HPW11}
F.~{Herwig}, M.~{Pignatari}, P.R. {Woodward}, D.H. {Porter}, G.~{Rockefeller},
  C.L. {Fryer}, M.~{Bennett}, R.~{Hirschi}, Ap. J. \textbf{727}, 89 (2011)

\bibitem{GZY13}
D.A. {Garc{\'{\i}}a-Hern{\'a}ndez}, O.~{Zamora}, A.~{Yag{\"u}e},
  S.~{Uttenthaler}, A.I. {Karakas}, M.~{Lugaro}, P.~{Ventura}, D.L. {Lambert},
  A\&A. \textbf{555}, L3 (2013)

\bibitem{Lod03}
K.~{Lodders}, Meteoritics and Planetary Science Supplement \textbf{38}, 5272
  (2003)

\bibitem{TMP07}
F.K. {Thielemann}, D.~{Mocelj}, I.~{Panov}, E.~{Kolbe}, T.~{Rauscher}, K.L.
  {Kratz}, K.~{Farouqi}, B.~{Pfeiffer}, G.~{Martinez-Pinedo}, A.~{Kelic}
  et~al., International Journal of Modern Physics E \textbf{16}, 1149 (2007)

\bibitem{ArG03}
M.~{Arnould}, S.~{Goriely}, Physics Reports \textbf{384}, 1 (2003)

\bibitem{KGB11}
F.~{K{\"a}ppeler}, R.~{Gallino}, S.~{Bisterzo}, W.~{Aoki}, Reviews of Modern
  Physics \textbf{83}, 157 (2011)

\bibitem{LLA14}
C.~{Langer}, O.~{Lepyoshkina}, Y.~{Aksyutina}, T.~{Aumann}, S.B. {Novo},
  J.~{Benlliure}, K.~{Boretzky}, M.~{Chartier}, D.~{Cortina}, U.D. {Pramanik}
  et~al., Phys. Rev. C \textbf{89}, 035806 (2014)

\bibitem{AAA14}
S.G. {Altstadt}, T.~{Adachi}, Y.~{Aksyutina}, J.~{Alcantara}, H.~{Alvarez-Pol},
  N.~{Ashwood}, L.~{Atar}, T.~{Aumann}, V.~{Avdeichikov}, M.~{Barr} et~al.,
  Nuclear Data Sheets \textbf{120}, 197 (2014)

\bibitem{GAA14}
K.~{G{\"o}bel}, P.~{Adrich}, S.~{Altstadt}, H.~{Alvarez-Pol}, F.~{Aksouh},
  T.~{Aumann}, M.~{Babilon}, K.~{Behr}, J.~{Benlliure}, T.~{Berg} et~al.,
  Journal of Physics Conference Series \textbf{in press} (2014)

\bibitem{CHO11}
M.B. {Chadwick}, M.~{Herman}, P.~{Oblo{\v z}insk{\'y}}, M.E. {Dunn},
  Y.~{Danon}, A.C. {Kahler}, D.L. {Smith}, B.~{Pritychenko}, G.~{Arbanas},
  R.~{Arcilla} et~al., Nuclear Data Sheets \textbf{112}, 2887 (2011)

\bibitem{ReL14}
R.~Reifarth, Y.A. Litvinov, Phys. Rev. ST Accel. Beams \textbf{17}, 014701
  (2014)

\bibitem{HIK08}
T.~{Hayakawa}, N.~{Iwamoto}, T.~{Kajino}, T.~{Shizuma}, H.~{Umeda},
  K.~{Nomoto}, Ap. J. \textbf{685}, 1089 (2008)

\bibitem{HiN00}
W.~{Hillebrandt}, J.C. {Niemeyer}, Annual Review of Astronomy and Astrophysics
  \textbf{38}, 191 (2000)

\bibitem{TRG11}
C.~{Travaglio}, F.K. {R{\"o}pke}, R.~{Gallino}, W.~{Hillebrandt}, Ap. J.
  \textbf{739}, 93 (2011)

\bibitem{BTG14}
S.~{Bisterzo}, C.~{Travaglio}, R.~{Gallino}, M.~{Wiescher}, F.~{K{\"a}ppeler},
  Ap. J. \textbf{787}, 10 (2014)

\bibitem{PiH12}
M.~{Pignatari}, F.~{Herwig}, Nuclear Physics News \textbf{22}, 18 (2012)

\bibitem{LHL03}
M.~Lugaro, F.~Herwig, J.C. Lattanzio, R.~Gallino, O.~Straniero, Ap. J.
  \textbf{586}, 1305 (2003)

\bibitem{TGA04}
C.~Travaglio, R.~Gallino, E.~Arnone, J.~Cowan, F.~Jordan, C.~Sneden, Ap. J.
  \textbf{601}, 864 (2004)

\bibitem{HKU07}
M.~Heil, F.~K{\"a}ppeler, E.~Uberseder, R.~Gallino, M.~Pignatari, Prog. Nucl.
  Part. Phys. \textbf{59}, 174 (2007)

\bibitem{RKV04}
R.~{Reifarth}, F.~{K{\"a}ppeler}, F.~{Voss}, K.~{Wisshak}, R.~{Gallino},
  M.~{Pignatari}, O.~{Straniero}, Ap. J. \textbf{614}, 363 (2004)

\bibitem{ZAB10}
Q.~Zhong, T.~Aumann, S.~Bishop, K.~Blaum, K.~Boretzky, F.~Bosch,
  H.~Br{\"a}uning, C.~Brandau, T.~Davinson, I.~Dillmann et~al., Journal of
  Physics: Conference Series \textbf{202}, 012011 (2010)

\bibitem{LBB13}
Y.A. {Litvinov}, S.~{Bishop}, K.~{Blaum}, F.~{Bosch}, C.~{Brandau}, L.X.
  {Chen}, I.~{Dillmann}, P.~{Egelhof}, H.~{Geissel}, R.E. {Grisenti} et~al.,
  Nuclear Instruments and Methods in Physics Research B \textbf{317}, 603
  (2013)

\bibitem{GBB10}
M.~Grieser, R.~Bastert, K.~Blaum, H.~Buhr, D.~Fischer, F.~Laux, R.~Repnow,
  T.~Sieber, A.~Wolf, R.~von Hahn et~al., \emph{The Diagnostics System at the
  Cryogenic Storage Ring CSR}, in \emph{IPAC2010} (Kyoto, Japan,
  http://accelconf.web.cern.ch/AccelConf/IPAC10/, 2010), Vol. MOPD092, p. 918

\bibitem{NHL11}
F.~Nolden, P.~H{\"u}lsmann, Y.~Litvinov, P.~Moritz, C.~Peschke, P.~Petri,
  M.~Sanjari, M.~Steck, H.~Weick, J.~Wu et~al., Nucl. Instr. Meth A
  \textbf{659}, 69 (2011)

\bibitem{VGN13}
W.~Vodel, R.~Geithner, R.~Neubert, P.~Seidel, K.K. Knaack, K.~Wittenburg,
  A.~Peters, H.~Reeg, M.~Schwickert, \emph{20 Years of Development of
  SQUID-based Cryogenic Current Comparator for Beam Diagnostics}, in
  \emph{IPAC2013} (Shanghai, China,
  http://accelconf.web.cern.ch/accelconf/IPAC2013/, 2013), Vol. MOPME013, p.
  497

\bibitem{SCL13}
D.~Shubina, R.B. Cakirli, Y.A. Litvinov, K.~Blaum, C.~Brandau, F.~Bosch, J.J.
  Carroll, R.F. Casten, D.M. Cullen, I.J. Cullen et~al., Phys. Rev. C
  \textbf{88}, 024310 (2013)

\bibitem{GLR12}
M.~{Grieser}, Y.A. {Litvinov}, R.~{Raabe}, K.~{Blaum}, Y.~{Blumenfeld}, P.A.
  {Butler}, F.~{Wenander}, P.J. {Woods}, M.~{Aliotta}, A.~{Andreyev} et~al.,
  European Physical Journal Special Topics \textbf{207}, 1 (2012)

\bibitem{SKL07}
K.H. Schmidt, A.~Keli\ifmmode~\acute{c}\else \'{c}\fi{},
  S.~Luki\ifmmode~\acute{c}\else \'{c}\fi{}, M.V. Ricciardi, M.~Veselsky, Phys.
  Rev. ST Accel. Beams \textbf{10}, 014701 (2007)

\end{thebibliography}
\end{document}